\documentclass[%
 aip,
 sd,%
 amsmath,amssymb,
 reprint,%
]{revtex4-1}

\usepackage{graphicx}
\usepackage{dcolumn}
\usepackage{bm}

\begin{document}

\preprint{AIP/123-QED}

\title{The geometro-hydrodynamical formalism  of quantum spinning particle}
\author{Mariya Iv. Trukhanova}
\email{mar-tiv@yandex.ru}

\affiliation{Faculty of physics, Lomonosov Moscow State University, Moscow, Russian Federation.}

\date{\today}%

\begin{abstract}
We present  the development of the realistic geometro-hydrodynamical formalism of quantum mechanics for the spinning particle, that involves the vortical flows and is  based on the idea, that the spinor wave represents a new type of physical field, propagating in space-time and influencing  the corpuscule embedded in the wave. We assume that this new field created by the intrinsic spin of the particle and can produce new intrinsic forces and torques affecting the particle with spin. We derive and discuss physical effects of this field. \end{abstract}

\maketitle

           \section{Introduction}
The gravity theories with torsion, where torsion is the independent characteristic of the space-time, attracted attention for a long time \cite{00}, \cite{001}. Several mathematical formulations of gravity with torsion were developed.  Cartan's torsion, also known as classical torsion  \cite{0}, is defined as the part of the affine connection of space-time, which is antisymmetric in its inferior indices. Affine connection is a basic physical feature that gives mathematical definition of the structure of space-time. For the first time E. Cartan have proved that even if the affine connection is not a tensor, its antisymmetric part should be a tensor. In this approach
we can represent the affine connection of space-time with torsion $\Delta^{\alpha}_{\ \mu\nu}=\Gamma^{\alpha}_{\mu\nu}-K^{\ \  \alpha}_{\mu\nu}$ as a sum of the Christoffel symbols (a particular case of the affine connection)
\begin{equation}\label{Kris}\Gamma^{\alpha}_{\mu\nu}=\frac{1}{2}g^{\alpha\beta}(\partial_{\nu}g_{\mu\beta}+\partial_{\mu}g_{\nu\beta}-\partial_{\beta}g_{\mu\nu})
                \end{equation}
symmetric in its two lower  indices, where $g_{\alpha\beta}$ - is the metric tensor $(g^{\alpha\beta}g_{\alpha\gamma}=\delta^{\beta}_{\gamma})$, and of the contortion tensor  \cite{01}

                 \begin{equation}K^{\ \ \alpha}_{\mu\nu}=-C^{\ \ \alpha}_{\mu\nu}+C^{\ \alpha}_{\nu \ \mu}-C^{\alpha}_{\ \mu\nu},
                \end{equation}    where $C_{\alpha\beta\gamma}$ is called the  Cartan's torsion tensor, which is antisymmetric  in the last two indices
                           \begin{equation}C^{\ \ \alpha}_{\mu\nu}=\frac{1}{2}(\Delta^{\alpha}_{\mu\nu}-\Delta^{\alpha}_{\nu\mu}).
                \end{equation}

                The manifold equipped with this connection was called a  $"$Riemann-Cartan  space$"$.
                The  affine connection different from Christoffel symbol characterizes  the geometry is not completely
described by the metric, but also a  independent characteristic - tensor \cite{011}.
The metric determines the structure of space-time, and in case of the curved space-time it differs from the metric of the flat Minkowski space-time.

On the other hand, as is well known, the torsion has $24$ independent components and  can be decomposed into three irreducible components: the torsion vector $C_{\beta}=g_{\mu\nu}C^{\mu\nu}_{\beta}$, the torsion  pseudo or axial vector $A_{\beta}=\varepsilon_{\beta\alpha\mu\nu}C^{\alpha\mu\nu}$ and the tensor $b_{\alpha\mu\nu}$. Thus the torsion is represented as

                      \begin{equation}\label{dec} C_{\alpha\mu\nu}=C_{\alpha}g_{\mu\nu}-C_{\mu}g_{\alpha\nu}
                      +\varepsilon_{\alpha\mu\nu\gamma}A^{\gamma}+b_{\alpha\mu\nu}.
                \end{equation}

At length, theoretical models, directed on the correct physical interpretation of the torsion, have emerged. Anyway, the torsion tensor undoubtedly represents a novel physical field, and the effects of its interaction with the matter can be predicted. Similarly to such fundamental fields like electromagnetic, which is generated by charge, and gravitation generated by mass, the torsion field should also have its own source. Some theoretical models assign this role to the spin \cite{0111}. This common interpretation  leads us to the idea that the spin creates a field which changes the structure of space-time. We regard  the torsion as not just a mathematical construction but a representation of some physical reality. In the   geometrical interpretation, the torsion field must be physical demonstration of the change in the geometry of space-time, the particle in the torsion field must interact with the structure of space-time.   In the paragraph below we briefly describe some of the models that have been built to provide correct interpretations of the torsion field and are significant in the context of our work.

In ref. \cite{02} a formalism was presented which makes it possible for a modified form of local gauge invariance and minimal coupling to be compatible with torsion. The torsion is assumed to be determined by the gradient of a scalar function or potential. In the context of this idea the torsion is derived from the potential.  The dynamic theory for this potential was obtained.
Another more general physical formulation of the torsion has been developed using decomposition of torsion into three irreducible components \cite{011} and representation of torsion via the vector and axial vector decomposition (\ref{dec}). The interaction of free matter fields with external torsion field had been discussed in set of Refs. \cite{1} - \cite{4}. This mathematical approach leads to several interesting physical consequences. The authors of refs. \cite{1}, \cite{2} and \cite{21} discuss the theoretical basis for the search of the possible experimental manifestations
of the torsion field at low energies.  The Pauli-like equation was derived and contains new torsion-dependent
terms which have a different structure as compared with the standard electromagnetic ones. In Ref. \cite{2}, in particular,  the author has studied the behavior of spinning and spinless particles in a space-time with torsion, and discussed possible physical effects generated by the background torsion.

Another theoretical research of general relativity with torsion is reviewed in Ref. \cite{40}. The Dirac equation in curved space with torsion is derived and the relationship between string theory and torsion is presented. In the reference \cite{41} the author calculated the probability of spin flip of an electron due to a torsion wave. It had been compared to
the electromagnetic case, and ways to detect torsion were discussed.

On the other hand,  a group of de Broglie's followers -  T. Takabayasi, J. Vigier and P. Holland \cite{T1} - \cite{T6} attempted to derive feasible model of the quantum matter applicable to subsequent  formulation of the quantum mechanical theory of a non-relativistic spinning particle. It was shown   that  this model can be constructed in a quite realistic manner, i.e., the authors have formulated the theory in a form of a specified hydrodynamics of a fluid carrying intrinsic angular momentum to be called "spin".  This new geometro-hydrodynamical interpretation connects the hydrodynamical formalism of the quantum mechanics for a spinning particle and the triad structure underlying the classical spin. This formalism leads to construction of a realistic physical model of the quantum matter. The main idea of new interpretation is that the spinor wave represents a new physical field, which exerts an influence on a corpuscle moving within it.  The  geometrical representation of a two-component spinor with an orthogonal set of coordinate axes - triad was originally developed by Kramer \cite{T1}. And the geometro-hydrodynamical representation of a Pauli field as an assembly of very small rotating particles continuously distributed in space was later evolved in Ref. \cite{T2} using Kramer's idea. This new consideration describes the Pauli field as a fictitious fluid of small particles bearing intrinsic angular momentum or spin, and introduces a triad structure underlying the classical spin \cite{T3},    where each fluid element is represented by this triad and the spin vector is fixed to the third axis of the triad. Based on this, the spinor is interpreted as defining a state of rotation \cite{T6}.

\section{General theory.}

\subsection{Tetrads and anholonomies.}
Long before the above articles the theory of gravity with torsion based on the autoparallelism geometry     had been developed in Ref. \cite{6}, \cite{61}.
In Ref. \cite{5} Cartan hypothesized that  "rotation of the matter leads to the torsion of space". He, however, did not specify there, which of the two torsions describes the torsion field generated by rotating matter. Einstein have restricted his description of the accelerated reference frames to the translational relativity. To describe all degrees of freedom in the accelerated reference frame one, however, should introduce a manifold of the local anholonomity rotation coordinates, regarded to as elements of the event space, in order to maintain rotational relativity of physical equations \cite{6}.
M. Carmeli has attempted \cite{7} to generalize the relativistic theory  incorporating anholonomic rotation coordinates into the structure of the events space. He has formulated principles of the rotational relativity, but did not specify the events space, which cannot be Minkowski time-space.
The main purpose of this article is to construct a realistic physical model of a quantum matter using the geometro-hydrodynamical formalism.
The main idea that was used in the hydrodynamical representation of the spinor wave is that this field represents a new type of physical field \cite{T6}, propagating in space and exerting its influence on a corpuscle of mass $m$ moving with it and embedded in the wave.     In this paper we are trying to combine the ideas developed by the abovementioned de Broglie's followers with the concept of a torsion field as a novel physical field generated by the spin.
We  characterize the spinor as defining a state of rotation using the triad model, - the system of orthonormal vectors oriented with respect to a set of fixed space axis \cite{1}.  In other words the spinor determines a system of triad vectors which indicates a state of rotation. Our purpose is to evolve the hydrodynamical model of spinning fluid in the direction of geometro-hydrodynamical interpretation in which each element of this fluid is considered as a triad.
The simplest generalization of the three-dimensional Euclidian  geometry to the case of the manifold of oriented points is the geometry of absolute parallelism (autoparallelism) constructed on the  manifold which is represented as a vector bundle with a base formed by the manifold of the translational coordinates and a fibre specified at each point by the field of an orthogonal coordinate frame or triad $\mathbf{e}_{a}(\mathbf{x},t)$, which varies from point to point  depending on $\mathbf{x}$, where $a$ is the number of the reference vector, and has rotational degree of freedom. We  represent the realistic geometro-hydrodynamical structure of the Pauli field as a fluid constituted by continuous distribution of triads, the dynamics of each being determined by its location in the field. The dynamical evolution of this triad is represented by Euler's angles ($\varphi, \theta, \chi$) \cite{2}, which are the coordinates of the fiber.  We  follow the main concept  that the local spin density, is fixed on the triad vector, which can be formed from the Pauli spinor $\psi$

\begin{equation} \label{spin}
s^{i}=\frac{\hbar}{2}\psi^+\sigma^{i}\psi, \qquad  i=1,2,3. \end{equation}

The natural way is to choose the triad third axis in the direction of spin-polarization  $\Sigma_i$
                \begin{equation} \label{spin2}  \Sigma_i=e_i^{3}=\frac{\psi^+\sigma_{i}\psi}{\psi^+\psi}, \qquad   e_i^{3}=e_i^{1}\times e_i^{2},
 \end{equation}      where the system of vectors $e_i^{a}$ determines the state of rotation and has the components with respect to an orthogonal set of fixed space axes $\mathbf{c}^j$, which is represented by the rotation matrix $\lambda_{ab}$ in the terms of Euler angles
                                                           \begin{equation}   \mathbf{e}_a=\sum_{b} \lambda_{ab}\mathbf{c}^b,                     \end{equation}                                     where the orthogonal transformation matrix $\lambda_{ab}$ determines a local transformation in each point of space
                  \begin{widetext}
                    \begin{equation} \label{a}       \lambda_{ab}=\begin{pmatrix}   \cos\phi\cos\vartheta\cos\chi-\sin\phi\sin\chi & \sin\phi\cos\vartheta\cos\chi+\cos\phi\sin\chi & -\sin\vartheta\cos\chi \\  -\cos\phi\cos\vartheta\sin\chi-\sin\phi\cos\chi & -\sin\phi\cos\vartheta\sin\chi+\cos\phi\cos\chi & \sin\vartheta\sin\chi \\  \cos\phi\sin\vartheta & \sin\phi\sin\vartheta & \cos\vartheta \end{pmatrix}
                                                 \end{equation}          \end{widetext}

                                        The quantum mechanical quantities are converted  using the rotation operator, specified by a matrix $\hat{U}^{1/2}$, expressed in terms of Euler's angles

                                       \begin{equation} \label{D} \hat{U}=\hat{U}^{1/2}(\varphi,\theta,\chi)=e^{-\frac{i}{2}\varphi \hat{\sigma}_z}\cdot e^{-\frac{i}{2}\theta \hat{\sigma}_y}\cdot e^{-\frac{i}{2}\chi \hat{\sigma}_z} \end{equation}

                                       in terms of Euler's angles ($\varphi, \theta, \chi$) the rotation matrix has the following form
\begin{equation} \hat{U}^{1/2}(\varphi, \theta, \chi)=\begin{pmatrix} \cos{\frac{\theta}{2}}e^{-i(\varphi+\chi)/2} & -\sin{\frac{\theta}{2}}e^{i(\varphi-\chi)/2} \\ \sin{\frac{\theta}{2}}e^{-i(\varphi-\chi)/2} & \cos{\frac{\theta}{2}}e^{i(\varphi+\chi)/2} \end{pmatrix}  \end{equation}
    and the first column of the transformation matrix gives the Eulerian representation of the unit spinor function in the spin space. The spin functions form a covariant basis. When we rotate the reference frame, spin functions are transformed with the operators of rotation $\hat{U}$ using the equation $\psi'_{m'}=\sum_{m}\psi_m U_{mm'}$. The  operators in a rotating coordinate system may be derived using  the transformation matrix  $\hat{O}'=U\hat{O}U^{\hat{-1}}$.

                                       The unit spinor, which defines a state of rotation in the spin space has the following representation in  terms of the Euler angles
                                              \begin{equation}        \psi'=\begin{pmatrix}
                                              R\cos\frac{\theta}{2}e^{-\frac{i}{2}(\varphi+\chi)} \\ R\sin\frac{\theta}{2}e^{\frac{i}{2}(-\varphi+\chi)}   \end{pmatrix},
                                                 \end{equation} where $R$ is real amplitude.

                                       Further we use Euler's angles to describe rotation. A certain rotation of a reference frame may be represented as the single rotation by the angle $\omega$ around the axis $\mathbf{n}$. This description leads to the following expression for the operator of rotation
           \begin{equation} \hat{U}=e^{-i\omega\mathbf{n}\cdot\mathbf{S}}=\hat{I}\cos{\frac{\omega}{2}}-i(\mathbf{n}\cdot\mathbf{S})\sin{\frac{\omega}{2}}.  \end{equation}

Physical meaning of the geometro-hydrodynamical interpretation of the quantum mechanics being developed here requires us to move from three-dimensional points manifold of the Euclidian geometry to the three-dimensional manifold of oriented points, defined as points of space bearing the oriented reference simplex and regarded to as rotating solids of infinitely small size. The three-dimensional space in this case is represented as a vector bundle with a base formed by the manifold of the translational coordinates and a fiber specified at each point by the orthonormalized references or triads.

The first hypothesis underlying the geometro-hydrodynamical model of the quantum mechanics is that rotation of an object in the physical space affects its geometry, which takes the form of a three-dimensional bundle defined by internal rotational coordinates, or the Euler's angles. A rotational metric should therefore exist, in addition to translational metric, and be determined with infinitely small rotation around the instantaneous axis of the triad. Rotation of the triad, which makes impossible any description of an object movement that doesn't involve rotational coordinates, has its physical manifestation in the torsion field characterized by Ricci's torsion tensor \cite{6}.

The second concept is that the spinor wave characterizing the state of the spinning particle, represents a new type of physical field, the torsion field, which is caused by internal motion of an assembly of the triads. The field of triads is defined in the hydrodynamical representation as a fluid of a continuous distribution of very small rotating bodies. This spatial distribution is specified by the hydrodynamical density $\rho=\psi^+\psi$.   The torsion field  influences on the corpuscle moving via it through the coordinates of the fiber or set of Euler angles varying from point to point. The dynamical evolution of the torsion field reduces itself  to the movement of the center of mass of the ensemble of the rotating bodies. In the other words, the non-observable wave function assumes the geometrical representation and is characterized by real physical field which has the geometrical nature.

The new formalism of the spinor wave field developed in this article keeps to the realistic geometro-hydrodynamical representation based on the ideas of Takabayasi, Vigier and Holland, but is the natural extension of this hydrodynamical formalism of a spinning particle. We try to interpret the wave function in terms of real geometrical form. This new formalism is based on the hypothesis that the fiber space (in the non-relativistic case) is a form of matter existence. Movement of rotating objects in the physical space affects the geometry of the event space, namely, it changes rotational structure of the bundle. The orientation of the triad with respect to the fixed set of space axes  is defined by the space and time-dependent set of Euler angles $\phi(\mathbf{x},t), \vartheta(\mathbf{x},t), \chi(\mathbf{x},t).$   The system of vectors $\mathbf{e}_{a}$, which are defined at each point of space,  is orthonormal, determines a state of rotation $e^{a}_i\cdot e^j_{a}=\delta_i^j$ and exists as basic vectors defined  and translatable in the absolute sense to any point of space in any direction. The indices  $a, b, c$ - are the triad indices and  indices $i, j, k$ - are the coordinate indices. The tensor in the triad frame transform as follows

                     \begin{equation} A^a_b=e^a_ie^j_bA^i_j.
                                                            \end{equation}

Using the approach developed in \cite{5}, \cite{6}, we introduce the matrix  transformation  $\lambda_{ab}$ (\ref{a}) denoting the components of triad $e_i^{a}$ with respect to the space axes in standard orientation $\mathbf{e}_a=\lambda_{ab}\mathbf{c}^b$. The tensor of angular velocity of the rotation of the reference frame $\omega^{j}_{k}$ can be determined by the relation
                                  \begin{equation} \label{w}  \omega^j_k=e_{a}^j\cdot\frac{de^{a}_k}{dt}, \qquad \omega_{jk}=-\omega_{kj}.
                                                            \end{equation}

                                                            Now it is natural to introduce the main properties of absolute parallelism geometry (autoparallelism). The concept of the connection of geometry can be written as    \cite{71}

                                                              \begin{equation}\Delta^i_{\ jk}=\Gamma^i_{jk}+\Upsilon^i_{\ jk},\end{equation} and

\begin{equation}\label{affine}
    \Delta^{i}_{jk}=e^{i}_a\partial_k e_j^a,\end{equation}
                                                              where the Christoffel symbols (\ref{Kris}),  depend on the metric tensor $g_{ij}=\eta_{ab} e^{a}_ie^{b}_j$, where the Minkowski metric is denoted by $\eta_{ab}$  and $\Upsilon^i_{\ jk}$ depends on a torsion tensor $\Omega^{\ \ i}_{jk}$ in the coordinate indices

                                                 \begin{equation} \Upsilon^i_{\ jk}=-\Omega^{\ \ i}_{jk}+g^{if}
                                                 \biggl(g_{jm}\Omega^{\ \ m}_{fk}+g_{km}\Omega^{\ \ m}_{fj}\biggr),\end{equation}
                                           where
                                    \begin{equation} \label{W}  \Omega^{\ \ i}_{jk}=\frac{1}{2}e^i_{a}\biggl(\partial_je^{a}_k-\partial_ke^{a}_j\biggr),
                                                            \end{equation}
                               which can be characterized as the object of anholonomity \cite{10}. In  case of the flat affine space with torsion, the quantities $\Upsilon^i_{\ jk}$ represent the local spin connection of space and are referred to as the Ricci rotation coefficients for the basis $e^{a}_i$. A space of events has two metrics - the Riemann flat metric and the three-dimensional Killing-Cartan metric \cite{6} $d\nu^2=d\chi_{ij}d\chi^{ij},$ where the infinitesimal increments can be given by the vector $d\chi^i=d\chi\cdot e^i_{\chi}=\omega^idt$.  The relationship (\ref{w}) is of particular interest, as it presents analytical validation of Cartan's hypothesis introduced in \cite{5}, It says that the angular velocity of matter rotation should induce torsion in the absolute parallelism geometry $\Omega^{\ \ i}_{jk}$. Rotation of the matter generates torsion of the absolute parallelism space, not of the Riemann-Cartan space. Rotation modifies the structure of space and the torsion physics results from this picture.

                                                            The  parallel displacement of the triad relative to the connection $\Delta^i_{\ jk}$ equals zero identically $\partial_ke^{a}_j-\Delta^i_{\ jk}e^{a}_i=0$. From this definition the connection can be defined as $\Delta^i_{jk}=e^i_{a}\partial_ke^{a}_j$ and the Ricci rotation coefficients proportional to the covariant derivative with respect to the Christoffel symbols
                                                            $$\Upsilon^i_{\ jk}=e^i_{a}\nabla_k e^{a}_j.$$

                                                            As a result the angular velocity of the triad must have the geometrical form
                                                       \begin{equation} \label{New} \omega^i_j=\Upsilon^i_{\ jk}\frac{dx^k}{dt},
                                                                      \end{equation}      where
                                                                      co-moving angular velocity is given by $\omega^{a}=1/2\varepsilon^{a}_{bc}e_j^{b}e_j^c.$

                                                             The angular velocity $3$-vector $\omega^i$ can be found from the relations defining rigidly rotating Cartesian coordinates. Easy to derive is that, in the flat space,  the vector of the triad, which provides a covariant specification of a state of rotation, in the flat space satisfies the equation
                                                         \begin{equation}  \label{e} \frac{de^{a}_j}{dt}+\Upsilon^i_{\ jk}\frac{dx^k}{dt}e^{a}_i=0,
                                                                         \end{equation}
                                                             which is responsible for the temporal dynamics of the triad vectors.

  \subsection{Non-riemannian geometry in Serret-Frenet frame. The elastic analogy.}
We formulate the model in terms of an orthonormal triad  $e^{a}_j=\mathbf{e}^{a}$, which can describe the space spin chain, which is the analogy to the elastic flexible rod.    We consider  a space curve with unit tangent $\mathbf{t}$, normal $\mathbf{n}$ and a binormal $\mathbf{b}=\mathbf{t}\times\mathbf{n}$ defined at every point on the curve such that the triad $(\mathbf{t}, \mathbf{n}, \mathbf{b})$ represents an orthonormal, Frenet frame. The orthonormal triad can be defined as an element of the group of the proper rotation $SO(3)$.
The configuration space for spin is organized by the set of points spanned by the vector
             \begin{equation}  \mathbf{S}=\frac{\hbar}{2}\mathbf{e}^{3}=\frac{\hbar}{2}\mathbf{t},
                                                                         \end{equation}
                                                                         where the tangent vector has the definition
    \begin{equation}   t^i= \frac{\psi^+\sigma^i\psi}{\psi^+\psi},            \end{equation}                                                   where $\psi$ - is the spinor function in the spin space.
Here we take into account the characteristic dynamical property of a triad, that the angular momentum of rotation (spin) is fixed on the third axis and has the constant magnitude or, in other words, we identify the spin-polarization  vector $\mathbf{\Sigma}=\mathbf{e}^{3}$ of a particle with the tangent vector of the space curve $\mathbf{t}$. In the Euler coordinates the vector $\mathbf{t}$  has the form of
                    \begin{equation}    \mathbf{t}= \begin{pmatrix} \sin\vartheta\cos\phi  \\ \sin\vartheta\sin\phi \\ \cos\vartheta \end{pmatrix}
                                                             \end{equation}

We introduce the configuration of spin chain that is described by the curve producing the position of a point as a function of arc length $s$.   The well-known formulae of differential geometry determine the set of Frenet-Serret equations
                  \begin{equation}  \label{Frenet1} \frac{d\mathbf{t}}{ds}=k\mathbf{n},
                                                                         \end{equation}
                  \begin{equation}  \label{Frenet2} \frac{d\mathbf{n}}{ds}=-k\mathbf{t}+\tau\mathbf{b},
                                                                         \end{equation}
                  \begin{equation}  \label{Frenet3} \frac{d\mathbf{b}}{ds}=-\tau\mathbf{n},
                                                                         \end{equation}
  where $k$ is the curvature
  \begin{equation}        k=|\frac{d\mathbf{t}}{ds}|        \end{equation}
  and $\tau$ the torsion of the space curve.
  The orthogonal trihedral defines the space curve within the rigid rotation by the Darboux vector $\mathbf{w}=\tau\mathbf{t}+k\mathbf{b}$, because when we move along the curve, triad undergoes   rotation and this rotation is described by the angular velocity $\mathbf{w}$. The Frenet frame rotates with this angular velocity moving along the curve. In terms of angular velocity the equations (\ref{Frenet1}) - (\ref{Frenet3})  can be written as
                               \begin{equation}  \label{Frenet4} \frac{d\mathbf{t}}{ds}=\mathbf{w}\times\mathbf{t}(s), \qquad
                               \frac{d\mathbf{n}}{ds}=\mathbf{w}\times\mathbf{n}(s), \qquad \frac{d\mathbf{b}}{ds}=\mathbf{w}\times\mathbf{b}(s).
                                                                         \end{equation}

On the other hand the geodesic equation (\ref{e}) for the triad vectors can be rewrite in the form
\begin{equation}  \label{Frenet5} \frac{d\mathbf{e}^{3}}{ds}=-\Upsilon^{3}_{\ kb}\mathbf{e}^{b}\frac{dx^k}{ds}=k\mathbf{e}^{2},
                                                                         \end{equation}
                  \begin{equation}  \label{Frenet6} \frac{d\mathbf{e}^{2}}{ds}=-\Upsilon^{2}_{\ kb}\mathbf{e}^{b}\frac{dx^k}{ds}=-k\mathbf{e}^{3}+\tau\mathbf{e}^{1},
                                                                         \end{equation}
                  \begin{equation}  \label{Frenet7} \frac{d\mathbf{e}^{1}}{ds}=-\Upsilon^{1}_{\ kb}\mathbf{e}^{b}\frac{dx^k}{ds}=-\tau\mathbf{e}^{2},
                                                                         \end{equation}
                    where the projected connection coefficients have the form of

                      \begin{equation}  \label{Frenet8} \Upsilon^{3}_{\ k2}\frac{dx^k}{ds}=-\Upsilon^{2}_{\ k3}\frac{dx^k}{ds}=-k,\end{equation} \[  \Upsilon^{1}_{\ k2}\frac{dx^k}{ds}=-\Upsilon^{2}_{\ k1}\frac{dx^k}{ds}=\tau\]

                     The rotation of the polarization vector observed $\mathbf{\Sigma}$  has  purely geometrical effect due to the non-Euclidian geometry and  the path of the spin field defines the congruence of the geodesics of the space.

\subsection{The analogy with spin-half condensates. Hydrodynamic representation of geometric quantities.}

To move from pure geometrical description to the hydrodynamical description we should specify vector and scalar fields that characterize torsion field as an object of "real" physics. In this interpretation a triad is an element of space and constitutes the field of frames which vary from point to point $\mathbf{e}^{a}(\mathbf{x},t)$ via the spacetime-dependent set of  Euler angles and  we transform the interpretation of torsion field in such a way that to introduce the antisymmetric field tensor as a function of the triads field.

Following the description developed in Ref. \cite{8} the analog of $4$-vector potential of the triad fluid is
\begin{equation}\Upsilon_{\alpha}=(\Upsilon_{0},\Upsilon_{j}), \qquad  and \qquad i,j=1, 2, 3 \end{equation}
  here we introduce the scalar $\Upsilon_{0}$ and vector $\Upsilon_{j}$ potentials of the torsion field using the analogy with vector and scalar potentials appearing in the Maxwell's electrodynamics.  This representation results from the important method of the theoretical physics that interprets the torsion field in terms of the covariant derivative. In this paper, the Greek superscripts/subscripts are the coordinate indices and   stand for $0, 1, 2, 3.$   The covariant derivative is determined by definition
                  \begin{equation}   \label{Cov}
            \nabla_{\alpha}A^{\beta}=\partial_{\alpha}A^{\beta}+\Delta^{\beta}_{\ \alpha\gamma}A^{\gamma}. \qquad \alpha,\beta,\gamma = 0, 1, 2, 3
  \end{equation}

  In the general case the affine connection is not a tensor, but the quantity $\nabla_{\alpha}A^{\beta}$ is a tensor if the affine connection transforms in a special non-tensor way \cite{2}. In the context of the flat space with torsion, the affine connection is $\Delta^{\beta}_{\ \alpha\gamma}=\Upsilon^{\beta}_{\ \alpha\gamma}$, if the metric is flat and the curvature tensor is zero.
Let's investigate the flat space with non-zero torsion.    In this case the spin connection

                           \begin{equation} \Upsilon^{\beta}_{\ \alpha\gamma}=-\Omega^{\ \ \beta}_{\alpha\gamma}+g^{\beta\nu}
                           \biggl(g_{\alpha\mu}\Omega^{\ \ \mu}_{\nu\gamma}+g_{\gamma\mu}\Omega^{\ \ \mu}_{\nu\alpha}\biggr)\end{equation}                  is characterized by  the object of anholonomity \begin{equation}\Omega^{\ \ \beta}_{\alpha\gamma}=
                           \frac{1}{2}e^{\beta}_{b}\biggl(\partial_{\alpha}e^{b}_{\gamma}-\partial_{\gamma}e^{b}_{\alpha}\biggr).
  \end{equation}

                     In the anholonomic coordinates the connection has the form

                       \begin{equation} \Upsilon^{a}_{\ b\gamma}=e^a_{\beta}\Upsilon^{\beta}_{\ \alpha\gamma}e^{\alpha}_b.
  \end{equation}

        The fundamental equation of a spin-$\frac{1}{2}$ particle in a space-time with torsion is the Dirac equation
        \begin{equation} \label{Dirac}  \hbar\gamma^{\mu}\biggl(\partial_{\mu}-\Upsilon_{\mu}\biggr)\Psi+imc\Psi=0,  \end{equation} where $\Psi$ is a bi-spinor, c, $\hbar$ are the speed of light and Planck constant, and quantity  $\Upsilon_{\mu}$ has the form
                           \begin{equation} \Upsilon_{\mu}=\frac{1}{4}\gamma_{a}\gamma_{b}\Upsilon_{\mu}^{\ ab}, \end{equation}
                                 where the Clifford algebra and Pauli matrices satisfy
          \begin{equation}   \gamma^0= \begin{pmatrix} I & 0 \\ 0 & -I   \end{pmatrix}, \ \  \gamma^{b}= \begin{pmatrix} 0 & \sigma^b \\ -\sigma^b & 0   \end{pmatrix}, \ \  \alpha^{b}= \begin{pmatrix} 0 & \sigma^b \\ \sigma^b & 0   \end{pmatrix},  \end{equation}                                                                     $and$                      \begin{equation}   \sigma^1= \begin{pmatrix} 0 & 1 \\ 1 & 0   \end{pmatrix}, \ \  \sigma^{2}= \begin{pmatrix} 0 & -i \\ i & 0   \end{pmatrix}, \ \  \sigma^{3}= \begin{pmatrix} 1 & 0 \\ 0 & -1   \end{pmatrix}.  \end{equation}

       After simple derivations, the analog of gauge potential $\Upsilon_{j}$ becomes represented by the  matrices acting on the spin vector
                     \[  \Upsilon_{j}=ia_{j}=ia_{j}^b\sigma^b=ia_{j}^1\sigma^1+ia_{j}^2\sigma^2+ia_{j}^3\sigma^3, \]\begin{equation}\Upsilon_{0}=i\frac{a_{t}}{c},  \qquad j=1,2,3,
                                     \end{equation}               where $\sigma^b$ are the Pauli matrices and $a_{j}^b$ belongs to the adjoint representation of the rotation group. The field $\Upsilon^{\mu}$, in the general formulation, is a non - Abelian gauge field, and has the components which do not commute. The curvature of this field can be determined as
                                    \begin{equation} W^b_{\mu\nu}=\partial_{\mu}\Upsilon^b_{\nu}-\partial_{\nu}\Upsilon^b_{\mu}-i[\Upsilon_{\mu},\Upsilon_{\nu}], \ W_{\mu\nu}=W^b_{\mu\nu}\sigma^b.   \end{equation}

                         Using the representation for bi-spinor $\Psi=\begin{pmatrix}\psi \\ \chi\end{pmatrix}$       we obtain two sets of equations

                  \begin{equation}  \biggl(i\hbar\partial_t-ic\hbar \Upsilon_0-mc^2\biggr)\psi +ic\hbar\biggl(\sigma\cdot\nabla-\sigma\cdot\mathbf{\Upsilon}\biggr)\chi=0\end{equation}
                  \[     \biggl(-i\hbar\partial_t+ic\hbar \Upsilon_0-mc^2\biggr)\chi -ic\hbar\biggl(\sigma\cdot\nabla-\sigma\cdot\mathbf{\Upsilon}\biggr)\psi=0
                  \]

                  The torsion in this case can be described using the potential and acts like a field. The potentials $a_t$ and $\mathbf{a}$  encode the spin state.  In the Euler angles, determining the arbitrary rotation of the spin frame, they can be represented as

                         \begin{equation} a_{\mu}^1=\frac{1}{2}\biggl(\nabla_{\mu}\chi\cos\varphi-\nabla_{\mu}\theta\sin\varphi\biggr),    \end{equation}
                         \[  a_{\mu}^2=\frac{1}{2}\biggl(\nabla_{\mu}\chi\sin\varphi+\nabla_{\mu}\theta\cos\varphi\biggr),
                         \]
                         \[      a_{\mu}^3=\frac{1}{2}\biggl(\nabla_{\mu}\chi+\cos\theta\nabla_{\mu}\varphi\biggr), \qquad
                                     \]   in the adiabatic limit, the off-diagonal components of $a_{\mu}^b$ can be neglected.

                  In the semi-relativistic approximation the Dirac equation (\ref{Dirac})  passes into the Pauli equation for a spin-half particles
                         \begin{equation} \label{Pauli}  i\hbar\partial_t\psi+\hbar a_t^b\hat{\sigma}^b\psi=\frac{1}{2m}\biggl(\hat{\mathbf{p}}-\hbar\mathbf{a}^b\hat{\sigma}^b\biggr)^2\psi
                         -\frac{\hbar}{2}\hat{\sigma}\cdot\mathbf{T}\psi,  \end{equation} where   the gauge potentials $a_t$ and $\mathbf{a}$  act on the spin vector.

               \subsubsection{Adiabatic approximation.}

                         In the adiabatic approximation we suppose that the off-diagonal components of $a_{\mu}$ are equal to zero and keep only the diagonal (i.e., $\sigma^3$) components.
                         In this case, the gauge field $\Upsilon_{\mu}$ has the form of
                                     \begin{equation} \Upsilon_{\mu}= i\begin{pmatrix} \frac{1}{2}\biggl(\nabla_{\mu}\chi+\cos\theta\nabla_{\mu}\varphi\biggr) & 0  \\ 0  & - \frac{1}{2}\biggl(\nabla_{\mu}\chi+\cos\theta\nabla_{\mu}\varphi\biggr) \end{pmatrix},\end{equation}    the sign $\pm$ denote the spin-up and spin-down states.  The case of fully spin-polarized system  corresponds to selecting the field $\Upsilon^{ad}_{\mu\uparrow}=\frac{i}{2}\biggl(\nabla_{\mu}\chi+\cos\theta\nabla_{\mu}\varphi\biggr)$. Pauli - Schrodinger equation for the spin-$1/2$ particles motion in the adiabatic approximation and full-polarized spin system  has the form of
                            \begin{widetext}
                                 \begin{equation}  \label{Hh}
                              i\hbar\frac{\partial\psi}{\partial t}=\frac{1}{2m}\Biggl(-i\hbar\partial-\hbar \mathbf{a}^3\Biggr)^2\psi
                              +\hbar a_{t}^3\psi-\frac{\hbar}{2}\sigma\cdot \mathbf{T}\psi+\frac{m}{2}\Upsilon\psi
  \end{equation}                                                      \end{widetext}
where      $\hat{\sigma}^{b}$ are the Pauli matrices,  $\hbar$ is the Planck constant,  $m$ denotes the mass  of  particles, c  is the speed of light in vacuum.    In our geometrical representation, the triad, as the element of space, has rotational degrees of freedom and determines the fiber specified at each point of the manifold, where the torsion field component $(a_t,\mathbf{a})$ determines the rotational movement of the triad fluid (in the hydrodynamical representation).

The new potential $\Upsilon$, which appears in the Pauli equation   (\ref{Hh}),  determines the stiffness of the spinning fluid, being analogy with the magnetic stiffness in the spin half condensate.   After simple calculations, the antisymmetric torsion field tensor can be represented in the form of
                          \begin{widetext}  \begin{equation}   f_{\alpha\beta}=\frac{\hbar}{2m}\biggl(\partial_{\alpha}\mathbf{e}^{2}\partial_{\beta}\mathbf{e}^{1}-
                            \partial_{\beta}\mathbf{e}^{2}\partial_{\alpha}\mathbf{e}^{1}\biggr) =\frac{\hbar}{2m}\mathbf{e}^{3}\cdot(\partial_{\alpha}\mathbf{e}^{3}\times\partial_{\beta}\mathbf{e}^{3}), \end{equation}\end{widetext} where the time and space components of this tensor $f_{\alpha\beta}(\mathbf{x},t)$ represent two types of the torsion field, - $\mathbf{e}(\mathbf{x},t)=\mathbf{f}_{t}$ and vorticity $T^i(\mathbf{x},t)=\frac{1}{2}\varepsilon^{ijk}f_{jk}.$ The existence of spin-vorticity

                           \begin{equation} \label{T3} T_i=\frac{\hbar}{2m}\varepsilon_{ijk}\mathbf{e}^{3}\cdot\partial_j\mathbf{e}^{3}\times
                           \partial_k\mathbf{e}^{3}\end{equation} is obliged to the vorticity of flow with the inhomogeneity of the triad field, $div\mathbf{T}=0.$

 The spatial distribution of the ensemble of triads is given by

 \begin{equation}\rho(\mathbf{x},t)=\psi^+\psi.\end{equation}

 Differentiation of $\rho(\mathbf{x},t)$ with respect to time and application of the Pauli equation (\ref{Hh}) leads to the continuity equation
  \begin{equation}  \label{Continuity}
                           \partial_t\rho+\nabla\cdot(\rho\mathbf{v})=0\end{equation}

 The center of mass of the ensemble of the triads associated with the velocity field of the particle has the representation \cite{8}
          \begin{equation}\label{v}   \mathbf{v}=\frac{1}{m}\mathbf{p}+\frac{\hbar}{4m}\sum_{a}\varepsilon^{ijk}\nabla  e_{j}^{a}\cdot e_{i}^{a}e^{3}_{k}, \end{equation}
where the last term represents the part of velocity field via the rotational orientations of the continuously distributed triads, it is assumed that the torsion field dynamically evolves so the triad rotates and the center of mass moves.
Substitution of the wave function $\psi=r e^{i\vartheta/\hbar}$ in the definition of the basic hydrodynamical quantities leads to the definition for the velocity field $m\mathbf{v}=\nabla\vartheta-\frac{\hbar}{2}\mathbf{a}$ and the momentum balance equation for the spinning fluid
\begin{widetext}\begin{equation}\label{j3} \rho(\partial_{t}+\mathbf{v}\mathbf{\nabla})\mathbf{v}=\rho\mathbf{e}+\rho\mathbf{v}\times\mathbf{T}+\frac{\mathbf{s}}{m}\cdot\mathbf{\nabla}\mathbf{T}
+\frac{\hbar^2}{2m^2}\rho\nabla\biggl(\frac{\triangle\sqrt{\rho}}{\sqrt{\rho}}\biggr)-\frac{1}{2}\rho\nabla\Upsilon.
\end{equation}                \end{widetext}

The momentum balance equation for the particle in the torsion field has the structure similar to the hydrodynamic equations for the spinor condensate.  The first two terms at the right side of the equation (\ref{j3}) describe the interaction with the torsion field, where the first term represents the effect of effective  $"electric"$   field on the  density and the second term in the definition (\ref{j3}) represents the effective $"magnetic"$  field $\mathbf{T}$ arising from the non-uniform spin textures of the Pauli field \cite{8}.  The second   term  is  torsion-dependent and characterizes the effect of spin-vorticity on the moving particle.   The third term is the effect of non-uniform torsion field on the spin and       characterizes the spin-torsion coupling in the non-uniform spin-vorticity field.  The fourth term is the quantum Madelund potential, the quantum pressure, which generates the quantum force, and the last term represents the gradient of the spin stiffness of the fluid

                  \begin{equation}   \Upsilon=\frac{\hbar^2}{4m^2}\partial_i\mathbf{e}^{3}
                  \cdot\partial_i\mathbf{e}^{3} \end{equation}

This new forces act inside the fluid and arise from the inhomogeneity of spin distribution.  New torsion terms   have new structure and the model developed here demonstrates the connection between geometry with torsion and geometro - hydrodynamical formalism developed by Takabayasi \cite{T1} - \cite{T6}, in which the hydrodynamical motion represents the motion of a quantum-mechanical spinning particle in such a way that each element of hydrodynamical fluid moves like a spinning particle under the action of internal potential and magnetic field.    In the geometro-hydrodynamical representation developed in our article, the torsion field is associated  with the fluid constituted by continuously distributed triads defined in the realistic hydrodynamical representation.
The equation of motion for the spin vector of particle $\mathbf{S}$ in the space with torsion can be derived using the definition for the axial vector of spin density
 $s^i=\frac{\hbar}{2}\psi^+\sigma^{i}\psi$

                        \begin{equation}  \label{M4}\partial_{t}\mathbf{s}+\nabla_i\mathbf{J}^i_s= \mathbf{s}\times\mathbf{T},        \end{equation}                                   where the spin current is $\mathbf{J}^i_s=\mathbf{s}v^i,$ and the  term at the right side of the equation (\ref{M4})  represents the torsion torque caused by the interaction with the torsion field inside the fluid. This term is the spin-torsion coupling term and can be interpreted as the torque caused by the interaction with spin-vorticity $\mathbf{T}$.
    To close the equations set (\ref{Continuity}), (\ref{j3}) and  (\ref{M4}) we need to derive the equation for the spin-vorticity   evolution, which can be obtained in the general form
      \begin{equation}\label{T}   \frac{\partial\mathbf{T}}{\partial t}=\nabla\times(\mathbf{v}\times\mathbf{T}). \end{equation}

\subsection{The spin helix. The eigenwaves.}
Let's describe the helical configuration of the spin $\mathbf{t}$, using  a simple form of the hydrodynamical   equation of motion  for the velocity field (\ref{j3})

                          \begin{equation}\label{j5}   m(\partial_{t}+\mathbf{v}\mathbf{\nabla})\mathbf{v}=-\frac{\hbar^2}{4m}\partial_k\biggl(\nabla\mathbf{t}
                          \cdot\partial^k\mathbf{t}\biggr), \end{equation}
                          and      the equation of motion for the spinning fluid (\ref{M4}), similar to the  equation in Ref. \cite{T2}

                              \begin{equation}  \label{M5}(\partial_{t}+\mathbf{v}\mathbf{\nabla})\mathbf{t}= \frac{\hbar}{2m}\mathbf{t}\times\partial^2\mathbf{t}.
         \end{equation}

         We need to close the set of equations (\ref{j5}) and (\ref{M5}) using the condition of incompressibility $\nabla\cdot\mathbf{v}=0$ and the Mermin-Ho relation for the spin-vorticity vector $\mathbf{T}=\nabla\times\mathbf{v}$

                     \begin{equation}  \label{T2}\mathbf{T}=\nabla\times\mathbf{v}=\frac{\hbar}{2m}\varepsilon_{ijk}t_i\nabla t_j\times \nabla t_k.
         \end{equation}

         The stable local spin forms helical configuration \cite{13}
                    \begin{equation} \mathbf{t}_0=\mathbf{c}_z\cos\vartheta+\sin\vartheta\cos(qz-\omega_0t)\mathbf{c}_x+\sin\vartheta\sin(qz-\omega_0t)\mathbf{c}_y, \end{equation}    where the small perturbations of the helical structure have the form of $\mathbf{t}=\mathbf{t}_0+\varepsilon_1\mathbf{n}+\varepsilon_2\mathbf{b}$,
                                            the curvature and torsion are $k=\sin\vartheta,$  $\tau=\cos\vartheta$.
                                            In the linear approximation the vorticity vector and the corresponding velocity field have the form
                                            \begin{equation}  \mathbf{v}(\mathbf{k},\omega)=\frac{\hbar q}{2mk^2}\sin\vartheta\biggl(k_xk_{\parallel}\mathbf{c}_x+k_yk_{\parallel}\mathbf{c}_y-k^2_{\perp}\mathbf{c}_z\biggr)\varepsilon_2(\mathbf{k},\omega),
         \end{equation}     and
                                        \begin{equation}   \mathbf{T}(\mathbf{k},\omega)=\frac{i\hbar q}{2m}\biggl(k_x\mathbf{c}_x-k_y\mathbf{c}_y\biggr)\varepsilon_2(\mathbf{k},\omega),
         \end{equation}       where the wave vector has the decomposition $\mathbf{k}=k_x\mathbf{c}_x+k_y\mathbf{c}_y+k_{\parallel}\mathbf{c}_z,$ and $k^2_{\perp}=k_x^2+k_y^2.$
                                                 The dispersion of eigenwave, characterized by the velocity evolution equation, is
                                         \begin{equation} \label{w} \omega=\frac{\hbar}{2m}q\cos\vartheta k_{\parallel}=\omega_0\frac{k_{\parallel}}{q}, \ and \                    \omega_0=\frac{\hbar}{2m}q^2\tau.          \end{equation}

The  relation (\ref{w}) expresses the dispersion of
wave  that emerges as a result of spin dynamics and has some analogy with the dispersion relation, derived in the Ref. \cite{13}.

 \subsection{The torsion-hydrodynamical waves}

The influence of the external torsion field on the dynamics of spinning particles is not great, similarly to the influence of the magnetic field. But, evolution of spin in the torsion field leads to the existence of new physical effects, described below. We can expect the existence of new physical processes which are the consequence of spin and vorticity dynamics. Spin dynamics causes changes in the wave dispersion in magnetized plasma \cite{9} - \cite{12} as well as the existence of novel branches of dispersion in systems of such kind.

Let's to derive dispersion characteristics of eigenwaves in the systems of neutral and spinning particles.  To do that let's analyze small perturbations of physical variables $(\mathbf{v}, \mathbf{s}, \mathbf{T})$ from the stationary state $f=f_0+\delta f$.  The unperturbed torsion field represents  as $\textbf{T}_{0}=T_{\parallel}\textbf{z}$  with respect to the propagation direction determined by the wavenumber $\textbf{k}=k_{\parallel}\textbf{z}+k_{\perp}(\cos\varphi\textbf{x}+\sin\varphi\textbf{y}),$ where the spin density   is  $\textbf{s}_{0}=s_{\parallel}\textbf{z}$.

In this case if we assume that linear excitations $\delta f$ are proportional to $exp(-\omega t + \textbf{k}\textbf{r})$ a linearized set of equations (\ref{j3}) - (\ref{T}) gives us the dispersion equation                                \begin{figure} [htbp]
   \centering\begin{tabular}{c}
     \includegraphics [scale=0.50] {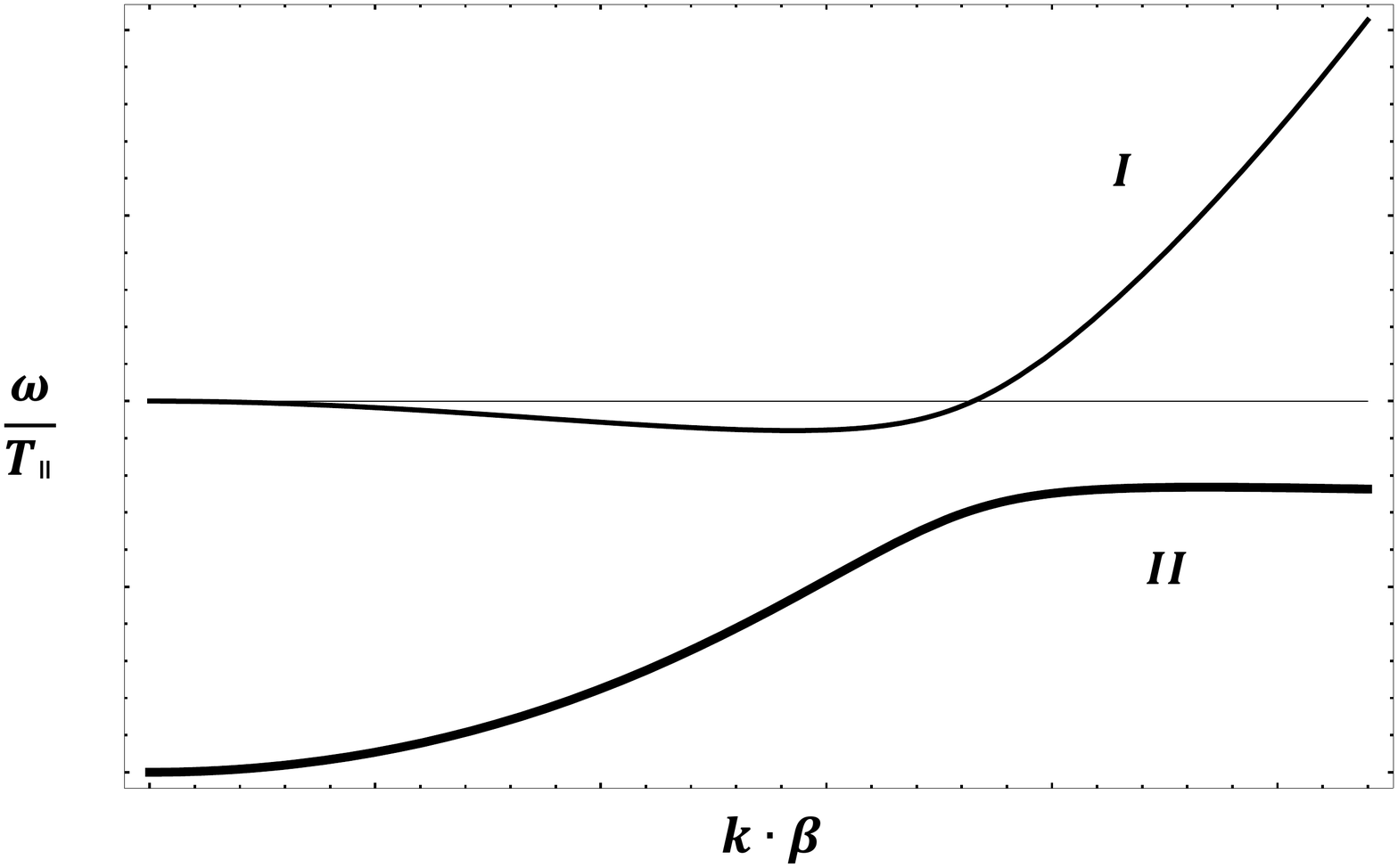} \\\end{tabular}
\caption{{\em The figure shows the dispersion characteristic of the torsion-hydrodynamical waves  (\ref{w3}). I - branch represents the solution with the sign $"+"$ and the II - branch the solution with the sign $"-"$.}
    }\label{p1} \end{figure}

\begin{equation} \label{w1} \omega^4-\omega^2\biggl(T_{\parallel}^2+\frac{\hbar^2}{4m^2}k^4-\frac{\hbar T_{\parallel}}{2m}k^2_{\bot}\biggr)
+\frac{\hbar^2k^2}{4m^2}T_{\parallel}^2k^2_{\parallel}=0.   \end{equation}

The equation (\ref{w1}) contains dispersion characteristic for
the torsion-hydrodynamical wave. In the quantum
case we consider this characteristic take on the form of
                                    \begin{widetext}
  \begin{equation} \label{w3}   \biggl(\frac{\omega}{T_{\parallel}}\biggr)^2=\frac{1}{2}
  \biggl(1+k^4\beta^4-k_{\bot}^2\beta^2\pm\sqrt{\biggl(1+k^4\beta^4-k_{\bot}^2\beta^2\biggr)^2
                                      -4\beta^4k^2\cdot k^2_{\parallel}}\biggr),\end{equation}  \end{widetext} where \[   \beta=\sqrt{\frac{\hbar}{2mT_{\parallel}}}.
                                              \]

The contribution of Madelung quantum potential
which is of purely quantum origin was taken into account
in the calculations. The term proportional to $\hbar^2$ characterizes the influence of the quantum pressure on the dynamics of wave. The term proportional to $\hbar$ is the torsion-dependent term and leads from the spin-torsion coupling in the non-uniform spin-vorticity field $T_{\parallel}$.

We consider propagation orthogonal to the external torsion field
                        \begin{equation} \label{w2} \omega^2=T_{\parallel}^2+\frac{\hbar^2}{4m^2}k_{\bot}^4-\frac{\hbar}{2m}T_{\parallel}k^2_{\bot}.                    \end{equation}

 As we can see, the wave orthogonal to the external torsion field $T_{\parallel}$ has the
property of stability.
 Waves that we
discovered possess the following feature: their frequencies
$\omega$ tends to $T_{\parallel}$ provided that wave number tends to zero $k_{\bot}\rightarrow 0.$

\section{Conclusions}
The theory developed in this article implements the new formalism of spinor field keeping to the realistic geometro-hydrodynamical representation. We developed the idea of Takabayasi, Holland and Vigier, that the spinor wave must represent a new physical field propagating in  space and influencing the corpuscle of mass moving within it.  We, at first step, went in the direction of geometrical description and introduced the manifold which is represented as a vector bundle with a base formed by the manifold of the translational coordinates and a fiber specified at each point by the field of an orthogonal coordinate frame or triad. In this non-relativistic interpretation, the triad becomes the element of space and we can identify the triad field as a fluid, and each element of this fluid has rotational degrees of freedom.

Characteristic dynamical property of each triad is that the spin vector of the particle is fixed on the third axis of triad vector. We identify the spinor wave with the field or fluid of frames and the motion of each frame defined by its location in the field. The spatial distribution of the fluid is determined by $\rho(\mathbf{x},t)=\psi^+\psi$ and the velocity field depends on the dynamical evolution of the triad field (\ref{v}). Using the interpretation of the torsion field in terms of vector and scalar potentials, we derived the momentum balance equation (\ref{j3}) and the spin density evolution equation (\ref{M4}), and we showed that the torsion field acts like a spin-vorticity field (\ref{T3}).  Dispersion branches of a novel type that occurs due to
spin-vorticity dynamics were predicted (\ref{w3}).

The expected physical effect of the torsion field, that follows from the system of equations (\ref{Frenet5}) - (\ref{Frenet8}), is rotation of the polarization vector of the particle in this field and, undoubtedly  rotation of the plane of polarization of the light moving through the external torsion field.

        \end{document}